\documentclass[%
 reprint,%
 amssymb, amsmath,%
 aip,cha,%
onecolumn]{revtex4-1}

\usepackage{floatrow}
\usepackage{etoolbox}
\usepackage{ulem}

\makeatletter
\patchcmd\captionlabel{\caption@setsubtype*{\FR@tmp}}{\caption@setsubtype*}{}{}
\makeatother

\newfloatcommand{capsidebox}{figure}[\capbeside\useFCwidth][\FBwidth]

\DeclareCaptionLabelFormat{rparen}{\noindent\llap{(\emph{#2})\,}}
\usepackage{adjustbox}

\usepackage{amsthm,amsmath,amssymb}
\usepackage{mathrsfs}

\usepackage {graphicx}
\usepackage{caption}
\usepackage{epsfig}
\usepackage{epstopdf}
\usepackage{subcaption}
\usepackage{amsmath,bm}
\usepackage{color}
\usepackage{lmodern}
\usepackage[T1]{fontenc}
\usepackage[utf8]{inputenc}
\usepackage{geometry}
\usepackage{subcaption}
\usepackage{docs}%
\usepackage{bm}%
\usepackage[colorlinks=true,linkcolor=blue]{hyperref}%
\usepackage{tabularx} 
\usepackage{booktabs} 
 
\expandafter\ifx\csname package@font\endcsname\relax\else
 \expandafter\expandafter
 \expandafter\usepackage
 \expandafter\expandafter
 \expandafter{\csname package@font\endcsname}%
\fi
\begin{document}

\title{Explore missing flow dynamics by physics-informed deep learning: the parameterised governing systems}

\author{Hui Xu}
\address{School of Aeronautics and Astronautics, Shanghai Jiao Tong University, Shanghai, 200240, China}

\author{Wei Zhang}
\email{waynezw0618@163.com}
\address{Science and Technology on Water Jet Propulsion Laboratory, Marine and Research Institute of China, Shanghai, China}

\author{Yong Wang}
\email{yong.wang@ds.mpg.de}
\address{Max Planck Institute for Dynamics and Self-Organization, 37077 G\"ottingen, Germany}

\date{}%
\revised{}%
\begin{abstract}
Gaining and understanding the flow dynamics have much importance in a wide range of disciplines, e.g. astrophysics, geophysics, biology, mechanical engineering and biomedical engineering. As a reliable way in practice, especially for turbulent flows, regional flow information such as velocity and its statistics, can be measured experimentally. Due to the poor fidelity or experimental limitations, some information may not be resolved in a region of interest. On the other hand, detailed flow features are described by the governing equations, e.g. the Navier-Stokes equations for viscous fluid, and can be resolved numerically, which is heavily dependent on the capability of either computing resources or modelling. Alternatively, we address this problem by employing the physics-informed deep learning, and treat the governing equations as a parameterised constraint to recover the missing flow dynamics. We demonstrate that with limited data, no matter from experiment or others, the flow dynamics in the region where the required data is missing or not measured, can be reconstructed with the parameterised governing equations. Meanwhile, a richer dataset, with spatial distribution of the control parameter (e.g. eddy viscosity of turbulence modellings), can be obtained. The method provided in this paper may shed light on data-driven scale-adaptive turbulent structure recovering and understanding of complex fluid physics, and can be extended to other parameterised governing systems beyond fluid mechanics.	
\end{abstract}
\maketitle

\section{Introduction}
In fluid mechanics, acquiring the flow data is one of the key tasks in science and engineering \cite{xiao2018,brunton2019}. Analysing or controlling the flow dynamics generally needs a rich data set or a continuous representation of the flow field. Conventionally, the flow dynamics is often conducted by either numerics or experiments. Numerically, the Navier-Stokes equations (NSEs), which govern the flow dynamics in a form of partial differential equations (PDEs), can be solved under typical assumptions or specific conditions within a typical range of precision. For high Reynolds number flows, or turbulent flows, directly solving the NSEs is extremely challenging due to limited computer resources and  large amounts of degrees of freedom \cite{moin1998,LAUNDER1974269,xiao2018}. To annihilate the restriction, various turbulence models, which are coupled with the NSEs, are proposed to save computational cost \cite{durbin2018}. Meanwhile, suspicion around correctness and applicability of modelling arises, which leads to unremitting pursuit of researches in turbulence modelling \cite{ling_kurzawski_templeton_2016,XIAO2016115,ling2016,Wang_2017}, although machine learning approaches have been used to enhance the predictive capability \cite{taghizadeh2020}.

In contrast, nowadays, gain of data from experiments is reliable, which generally has no restriction from the Reynolds number. However, due to the precision and uncertainty of experimental equipment  \cite{XIAO2016115}, measured data can be obtained in a finite set of locations, and are precise only in a limited-order statistical sense, which are not as sufficiently accurate as the solutions described by the NSEs. In a statistical sense, the experimental data can be generally regarded as an approximation of the solutions of the NSEs. It turns out that leveraging the limited experimental data to the underlying governing equations appears as a critical question in modelling and recovering the flow dynamics. In order to address this problem, we here consider the reformed governing equations where the control parameter is regarded as an unknown, and employ the physics-informed learning technique to preserve fundamental physical principles \cite{taghizadeh2020,Shukla2020,raissi2020HFM}, by which the parameterised governing equations and the missing flow dynamics can be physically determined at the same time. To achieve the above, the parameterised governing equations are treated as a physical constraint, which is used to form a loss function in the training of neural networks \cite{raissi2019}.

A data-driven discovery of PDEs, named PDE-FIND, was proposed by Rudy and his collaborators. Especially, for the NSEs, the PDE-FIND was used to identify the Reynolds number in the vorticity transport equation for fluid mechanics\cite{Kutz2017}. The research group of Karniadakis has made a great contribution in physics-informed neural networks \cite{raissi2019, Meng2020mpinn, Meng2020ppinn} and their applications to fluid flows. Generally, physics-informed neural networks can be used for both forward and inverse problems. For forward problem, Raissi et al. proposed the hidden fluid mechanics \cite{raissi2020HFM}, which provides a new way to learn velocity and pressure fields from flow visualizations. Most recently Jin et al. proposed the NSFNets to solve two different, velocity-pressure and vorticity-velocity, formulations of the incompressible NSEs, in the sense of direct numerical simulation for Kovasznay flow and turbulent channel flow \cite{JinNSFnets}. Other applications are for high speed inviscid flow \cite{Mao2020} and vortex-induced vibration \cite{Raissi2019viv}. For inverse problems, physics-informed neural networks were used to identify the Reynolds number from flow data for the low Reynolds number flow past a cylinder \cite{raissi2019}, and to discover universal variable-order fractional model for turbulent Couette flow \cite{Mehta2019}. 

Built on the above evidence, the generalisation of physics-informed neural networks forms the basis of the parameterised governing system in exploring missing flow physics and exploiting the essential of modelling parameters, specially for turbulent flows. To overcome numerical stiffness, a recently developed learning rate annealing algorithm \cite{Wang2020} is adopted in this work. The proposed idea can be extended to a broad range of applications where PDEs appear as physical constraints and the control parameter of the PDEs plays a role in dominating the physics.

\section{Methodology}

Mathematically, we consider the following system as a general description of physical phenomena in a region $\Omega$:
\begin{equation}\label{eq:1}
\mathcal{L}^t(\vec{p}; \vec{p}_0, \vec{g}_\Gamma, \nu)=\vec{f},
\end{equation}
where the operator $\mathcal{L}^t$ represents the evolution of $\vec{p}$, and $\vec{p}_0$ is the initial condition. $\vec{g}_\Gamma$ is the boundary conditions where $\Gamma$ denotes the boundary $\partial \Omega$ of $\Omega$. $\nu$ denotes the control parameter. For a steady system, the corresponding representation can be written as $\mathcal{L}^\circ(\vec{p}; \vec{g}_\Gamma, \nu)=\vec{f}$. In real-world applications, in the sense of temporal (or ensemble) {\it averaging} or spatial {\it filtering}, the control parameter can be represented as an effective one, $\nu_{\rm eff}$, which is the summation of the physical $\nu$ and the modelled $\nu_{t}$, while the original expression of the governing system is kept the same.  For instance,  in turbulence modelling, at any level of turbulence, closure of the averaged or filtered NSEs requires the parameter, eddy(subgrid) viscosity, which is introduced in large-eddy simulations (LES), scale-resolving simulations (SRS) and Reynolds-averaged Navier-Stokes (RANS) methods \cite{spalar2009,durbin2018}. Once the effective control parameter $\nu_{\rm eff}$ is introduced as an unknown parameter, $\mathcal{L}^t(\vec{p}; \vec{p}_0, \vec{g}_\Gamma, \nu_{\rm eff})=\vec{f}$ or $\mathcal{L}^\circ(\vec{p}; \vec{g}_\Gamma, \nu_{\rm eff})=\vec{f}$ can be regarded as the parameterised physical constraint which is used to explore missing flow dynamics under the framework of the physics-informed neural networks.

Let us consider the Leonardo's self-portrait, shown in Figure \ref{fig:1} (A), as a physical field. It is assumed that a part of Leonardo's hair is missing. In order to fix the portrait,  we need to find a pattern which is consistent with the missing one. Without referring the  potential governing system of the Leonardo's portrait, by learning the hair pattern beyond the information-missing region, it is possible to create a pattern which matches with the pattern around the region. It is clear that in principle, there should exist infinite possibilities. {\it So, by virtue of limited information, how to physically make it unique as possible as we can?} This is termed as ``{\it the parameterised physics-informed deep learning strategy}''.

\begin{center}
\includegraphics[height=2in,angle=0]{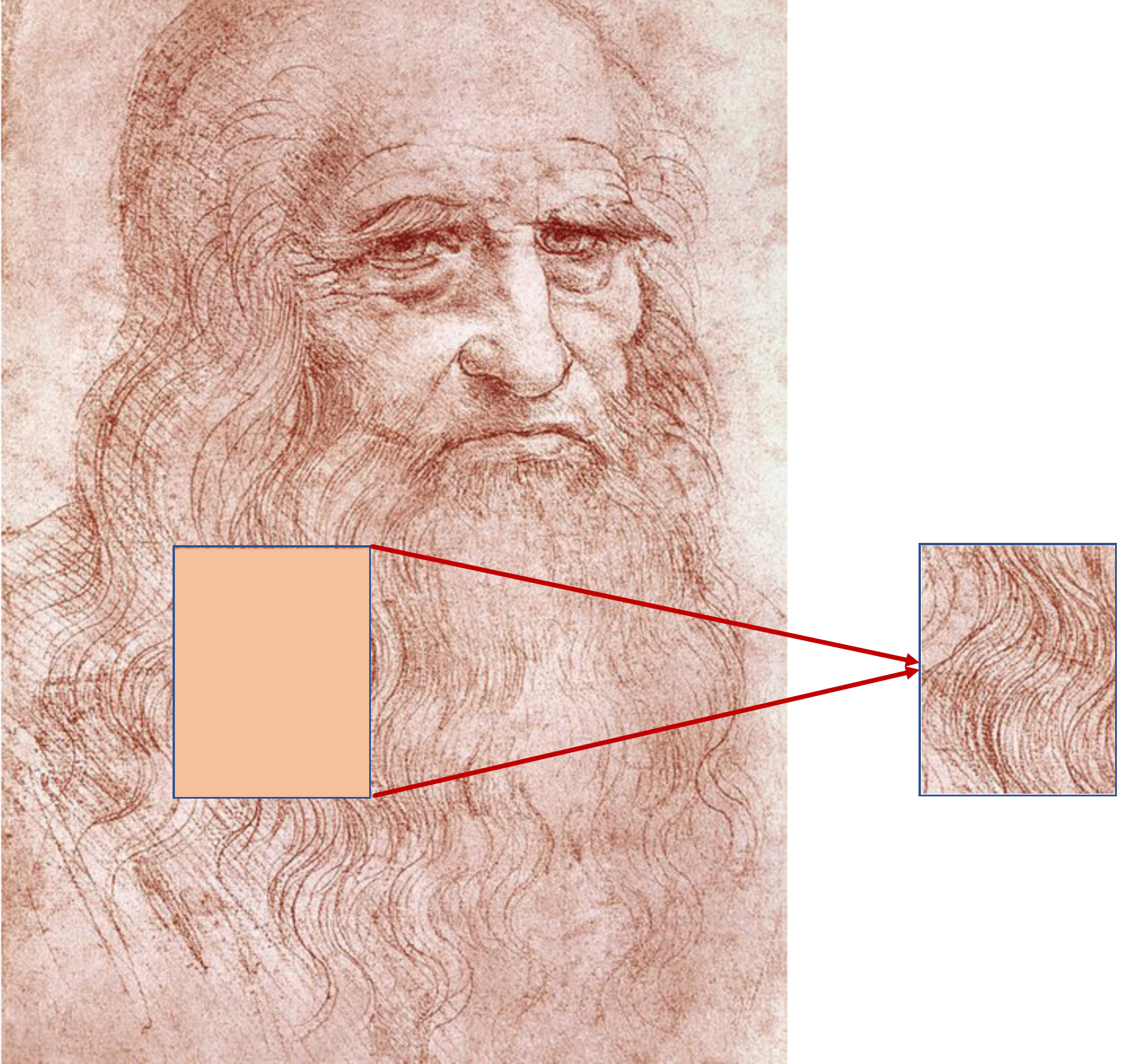} \hspace{1cm}
\includegraphics[height=1.5in,angle=0]{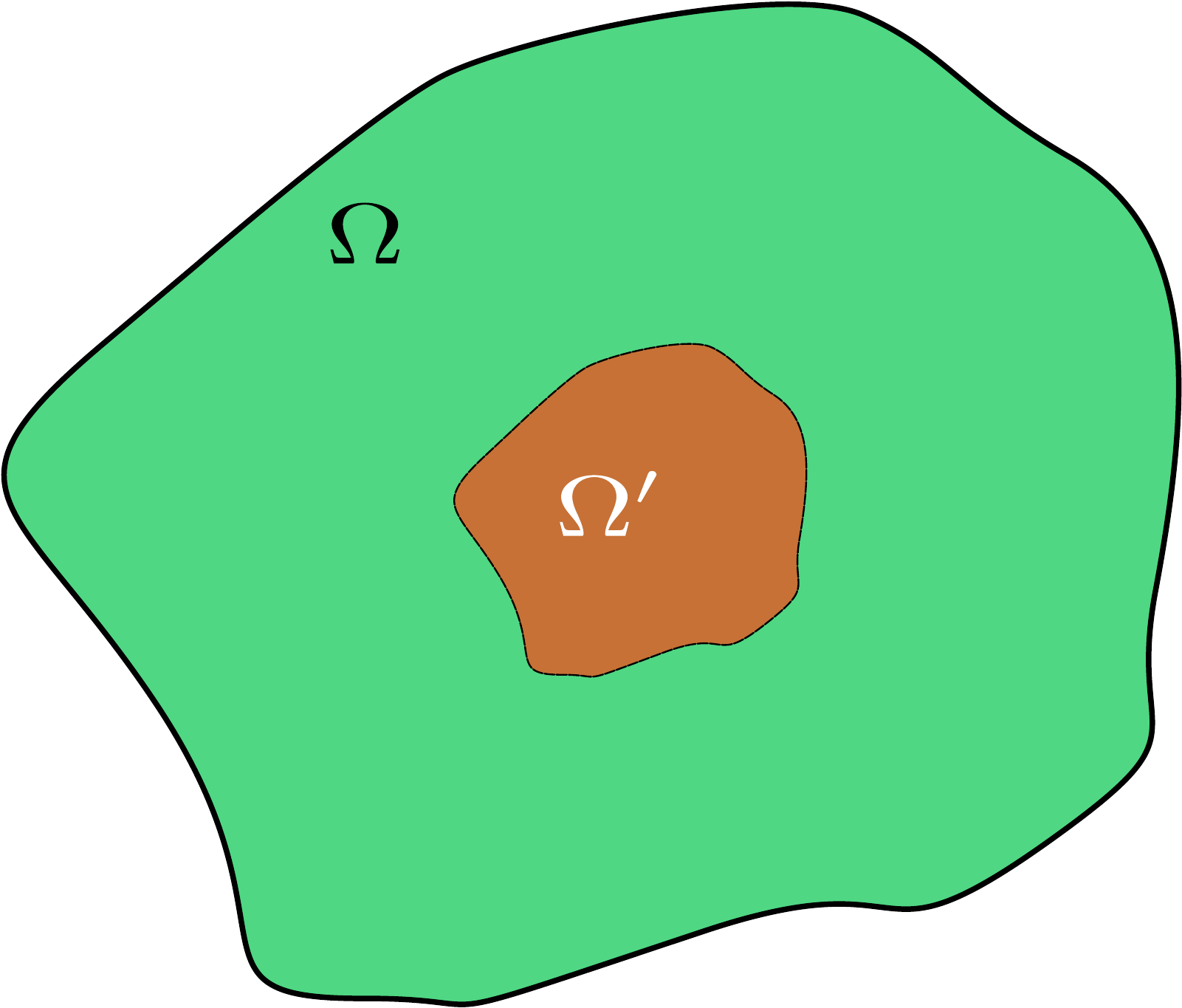} \\
\hspace{1cm} (A)  Leonardo's hair  \hspace{3cm} (B) Domain $\Omega$ and sub-domain $\Omega^\prime$ 
\captionof{figure}{Illustration of missing information: (A) Leonardo's self-portrait with a part of curly hair missing; [Generated from \url{https://en.wikipedia.org/wiki/Leonardo_da_Vinci}] (B) Mathematical representation of the domain $\Omega$ and the sub-domain $\Omega^\prime$ on which the information is missing.}
\label{fig:1}
\end{center}

As illustrated in Fig. \ref{fig:1}(B), $\vec{p}$ is known on $\Omega\setminus \Omega^\prime$ and unknown on $ \Omega^\prime$. Theoretically, in fluid mechanics, if the effective eddy viscosity $ \nu_{\rm eff}$ can be estimated or physically formulated,  the missing $\vec{p}$ on $ \Omega^\prime$ can be solved from Eq. (\ref{eq:1}). In the framework of the parameterised physics-informed learning, solving  Eq. (\ref{eq:1}) is avoided and Eq. (\ref{eq:1}) only leads us to directly find a physical unique solution on $ \Omega^\prime$ in terms of $\vec{p}$ on $\Omega\setminus \Omega^\prime$. Let us turn back to the physical world of the fluid dynamics which is governed by the PDEs, e.g. the incompressible NSEs as follows
\begin{equation}\label{eq:3}
\mathcal{L}^t(\{\mathbf{u},p\};\mathbf{g},\nu)\equiv\left\{\begin{array}{rl}
\displaystyle\frac{\partial \mathbf{u}}{\partial t}-\nu\nabla^2\mathbf{u}+\mathbf{u}\cdot\nabla\mathbf{u}+\nabla p&=\mathbf{f}, \mbox{on } \Omega\\[1em]
 \nabla\cdot\mathbf{u}&=0,\\[1em]
\mathbf{u}|_{\Gamma=\partial\Omega}& =\mathbf{g}.
\end{array}\right.
\end{equation}
Theoretically, Eq. (\ref{eq:3}) can be used for all kinds of incompressible flows. However, with extremely small viscosity $\nu$, and thus very high Reynolds number, solving Eq. (\ref{eq:3}) is extra challenging in practical applications. In another aspect, although some data can be obtained experimentally, it is also challenging, if possible, to establish the direct connection between Eq. (\ref{eq:3}) and the gained experimental data. In engineering applications, the RANS equations, which are derived by the Reynolds decomposition of flow fields, have been well validated for engineering applications by experimental data and direct simulations of the NSEs \cite{spalar2009,
durbin2018}. The key question in RANS is to model the Reynolds stress $\tau_{ij}$, which is represented as follows
\begin{equation}
\tau_{ij}=-\rho  \overline{u_i^\prime u_j^\prime},
\end{equation}
where $\overline{(\cdot)}$ denotes the Reynolds average or the spatial filtering, and $u^\prime_i=u_i-\overline{u}_{i}$. Similar to Fick's law,  $\rho  \overline{u_i^\prime u_j^\prime}$ can be modelled as
\begin{equation}
\rho  \overline{u_i^\prime u_j^\prime}=-\rho\nu_t (\frac{\partial \bar{u}_i}{\partial x_j} + \frac{\partial \bar{u}_j}{\partial x_i}),
\end{equation}
where $\nu_t$ denotes the eddy viscosity. Although it is named as viscosity, $\nu_t$ is an artificial parameter different from the molecular viscosity in the original equations. Similar idea has been adopted to the large eddy simulation (LES), where sub-grid eddy viscosity $\nu_{\rm sgs}$ is introduced to model the subgrid scale turbulent flow. For hybrid RANS/LES such as detach eddy simulation (DES) and scale adaptive simulation (SAS), $\nu_t$ in the control system has different meaning in different region of the computational domain, depending on the resolution of the turbulence scales. Normally, $\nu_t$ is flow dependent. In the turbulence community, various modeling of the parameter $\nu_t$
has been studied for several decades. With the big data and machine learning, model constants of the RANS equations are tuned by fitting \cite{ling_kurzawski_templeton_2016,poroseva2016,edeling2017,xiao2018}. However, compared with small data sets, observed discrepancies are not explained any better with big ones \cite{edeling2017}, and modelling $\nu_t$ so as the $\nu_{\rm sgs}$ in a universal sense still remains as a persistent challenging because of the case-by-case dependence. Luckily, engineering practices demonstrate that predictions of the RANS equations for complex flows are mostly quite satisfactory. It indicates that utilization of $\nu_t$ as an extra parameter can be a reliable way to establish the fitting between averaged/filtered experiment data and solutions of the corresponding control equations.  

As indicated in the pioneering work \cite{durbin2018}, representation of turbulent transport by an eddy viscosity underlies most of the related theoretical work. With this idea, the classical RANS equations can be rewritten as (without considering the auxiliary equations)
\begin{equation}\label{eq:4}
\mathcal{L}^t(\{\bar{\mathbf{u}},p\};g,\nu+\nu_{\eta})\equiv\left\{\begin{array}{rl}
\displaystyle\frac{\partial \bar{\mathbf{u}}}{\partial t}-(\nu+\nu_{\eta})\nabla^2\bar{\mathbf{u}}+\bar{\mathbf{u}}\cdot\nabla\bar{\mathbf{u}}+\nabla \bar{p}&=\mathbf{f}, \mbox{on } \Omega\\[1em]
 \nabla\cdot\bar{\mathbf{u}}&=0,\\[1em]
\bar{\mathbf{u}}|_{\Gamma=\partial\Omega}& =\mathbf{g},
\end{array}\right.
\end{equation}
where $\nu_{\rm \eta}$ is an artificial viscosity term introduced here for parameterisation similar to that described above. From the control parameter point of view, the data-based determination of $\nu_{\rm \eta}$ indicates  ``{\it scale adaption}'' of recovering turbulent structures, same as $\nu_{\rm t}$ in DES or SAS. The above equation is the system of the operator equation, Eq. (\ref{eq:1}), in a concrete way. For steady cases, the unsteady term $\partial \bar{\mathbf{u}}/{\partial t}$ is neglected.  It should be mentioned that reformulating the NSEs does not imply employment of the RANS approach since parameterisation of $\nu_{\rm eff}=\nu+\nu_{\rm \eta}$ appears only as an undetermined parameter. The parameterisation is significantly distinguished from the conventional tuning of model constants. The question remains as ``{\it how to determine the parameter $\nu_{\rm eff}$ (or $\nu_{\rm \eta}$) regarded as a spatial distributed variable?}''. 

With the paramaterised NSEs, assuming only limited data are available on a set $\mathcal{X}\in \Omega\setminus\Omega^\prime$ (cardinality $\mathcal{N}=|\mathcal{X}|$) of finite scattered measurement locations, we are interested in the missing flow dynamics on $\Omega^\prime$ where the specified point set $\mathcal{X}^\prime$ with cardinality $\mathcal{N}^\prime=|\mathcal{X}^\prime|$ is chosen to represent the dynamics. On $\mathcal{X}$, the set of measured averaged/filtered $\mathbf{u}$ is denoted by $\bar{\mathcal{S}}_{\mathcal{X}}$ and on $\Omega^\prime$, the set of missing averaged/filtered $\mathbf{u}$ is denoted by  $\bar{\mathcal{S}}_{\mathcal{X}^\prime}$. Further, the undetermined $\nu_{\rm eff}$ appears as a spatial variable defined on $\Omega$, and will be inferred later. Here, the target is to approximate the functional relation $(t,x,y,z)\mapsto (u,v,w,p,\nu_{\rm eff})$ by physics-informed neural network $(t,x,y,z)\mapsto (e_u,e_v,e_w,e_{\rm div})$. The parameterised NSEs are encoded in $e_u$, $e_v$, $e_w$, and $e_{\rm div}$, which represent the residuals of the momentum equations and  the continuity equation, respectively. To overcome numerical stiffness that lead to unbalanced back-propagated gradients during model training, a recently developed learning rate annealing algorithm is adopted \cite{Wang2020}. The physics-informed residuals $e_u$, $e_v$, $e_w$, and $e_{\rm div}$ are minimized with the mean squared error loss function on the time set $\mathcal{T}$
\begin{equation}
{\rm Loss}=\mathbb{E}_{\mathcal{T}}[{\rm MSE}_{\rm Data}+{\rm MSE}_{\rm PDEs}]
\end{equation}
where
\begin{equation}
{\rm MSE}_{\rm Data}=\mathbb{E}_{\mathcal{X}}\left[\|\mathbf{u}(t;\mathcal{X})-\mathbf{u}_{(t;\mathcal{X})}\|^2\right],
\end{equation}
and
\begin{equation}
{\rm MSE}_{\rm PDEs}=\mathbb{E}_{\hat{\mathcal{X}}}\left[e_u^2+e_v^2+e_w^2+e_{\rm div}^2\right].
\end{equation}
Here $\hat{\mathcal{X}}\in \Omega$. The operator $\mathbb{E}_\#[\cdot]$ means the average on the set `\#' in a distribution sense. $\mathbb{E}_\mathcal{T}[\cdot]$ is an identity operator for stationary modellings. Generally, the location cluster can be  generated in a probability sense. Once the above minimisation problem is solved, the extra variables $p$ and $\nu_{\rm eff}$ can be discovered. 

\section{Numerical configuration}
\begin{figure}[hbt!]
\begin{center}
\includegraphics[height=1.7in,angle=0]{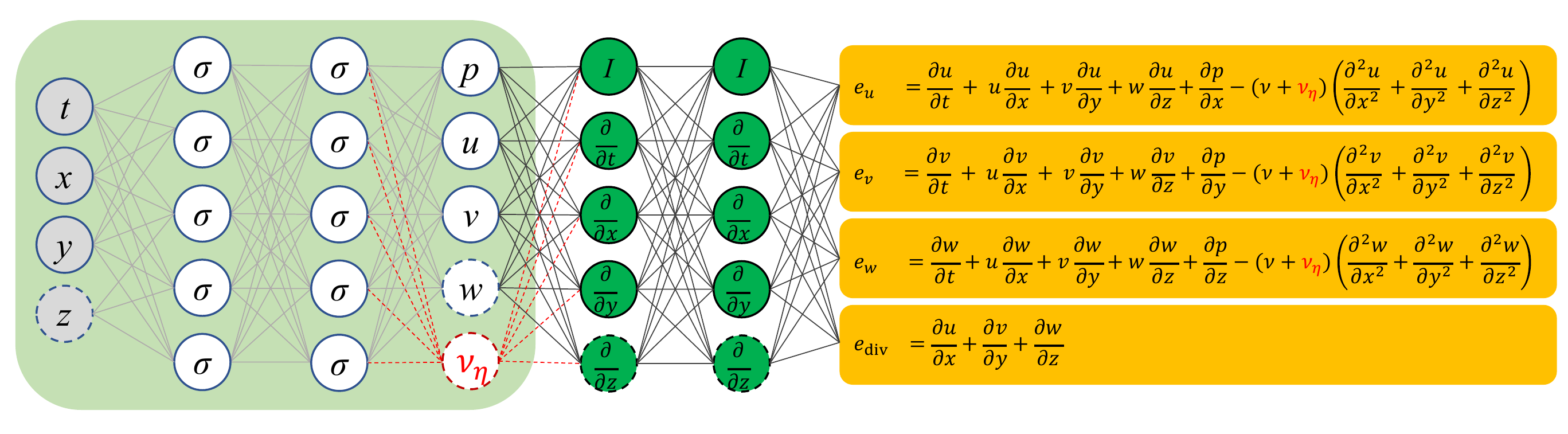} 
\captionof{figure}{Illustration of the physics-informed neural networks. The parameter $\nu_{\rm \eta}$ is regarded as an output variable of the neural network.}
\label{fig:nn}
\end{center}
\end{figure}

The diagram of the physics-informed neural network is illustrated in Fig. \ref{fig:nn}. We here treat the parameter $\nu_{\rm \eta}$ as an output of the deep neutral network. During the training process, the required weights of neural network are determined by the loss function formed by the residual of the governing equations. Once the training convergence is achieved, the variables $u$, $v$, $w$, $p$ and $\nu_{\rm \eta}$ can be inferred. The training datasets were generated by computational fluid dynamics (CFD) and the open-source software OpenFOAM \cite{openfoam1998} was used to implement required computations. In two-dimension, only values of the velocity components $u$ and $v$ at mesh nodes are considered as training data. 

To check the ability of the method, we considered three different cases, related to the parameterised governing equations of RANS, URANS and SAS for two different turbulent flows, respectively. SAS denotes ``{\it scale-adaptive simulation}'' \cite{spalar2009}, and typically allows simulation of unsteady flows with both RANS and LES content. It should be noted that only regional velocity is used as training data, while velocity, pressure and artificial viscosity in the entire domain are inferred. 


Two flows, including a steady turbulent flow over a back-ward facing step (RANS) and an unsteady turbulent flow past a cylinder (URANS and SAS), were involved. For the steady one, there are $1.2\times 10^4$ cell nodes in total, while the random sampling nodes are $3\times 10^3$ for training data (on $\Omega\setminus \Omega^\prime$) and $4.3\times 10^3$ constraints of the parameterised governing system (on $\Omega$), respectively. A neural network with 6 hidden layers and 50 neurons per layer is considered. $2\times10^5$ epochs are used. For the unsteady flow past a cylinder, there are 351 snapshots and $5.1\times 10^5$ cell nodes on every snapshot, thus $1.7901\times 10^8$ in total. The random sampling nodes are $1.2\times 10^7$ and $1.4\times 10^7$ respectively, on all time snapshots. Corresponding neural network has 10 hidden layers and 96 neurons per layer. The learning rate is gradually decreased from $1.0\times 10^{-3}$ to $1.0\times10^{-6}$ with 400 epochs at each stage. The Adam optimizer is adopted for all cases.

\section{Results}

\begin{figure}[hbt!]
\begin{center}
\includegraphics[height=1in,angle=0,width=0.8\linewidth]{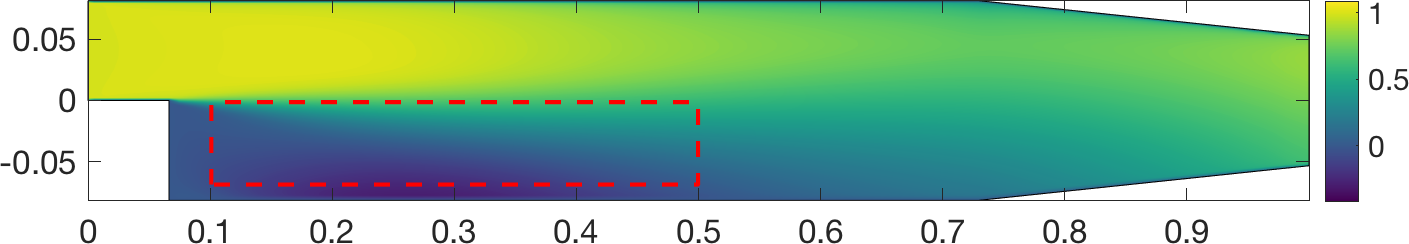} \\[1em]
\includegraphics[height=1in,angle=0,width=0.8\linewidth]{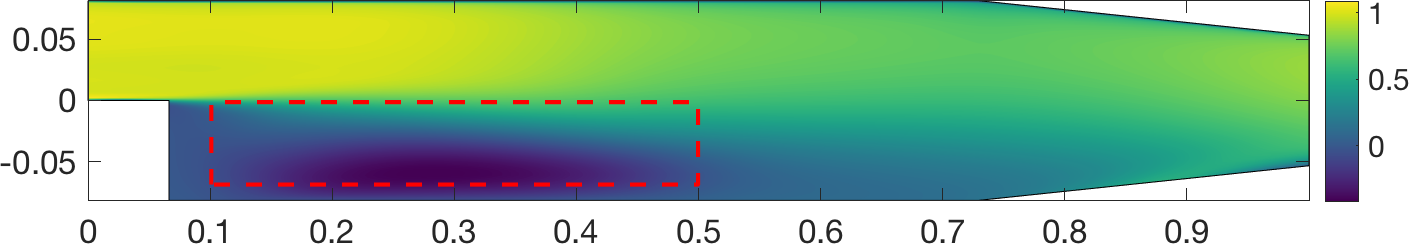} 
\captionof{figure}{Contours of the velocity component $u$ in the streamwise direction for the steady turbulent flow over a backward facing step. (Top) from the {\it SST} $k -\omega$ model; (Bottom) inferred by the physics-informed neural network. The dash-lined box indicates the region where the flow information is missing.}\label{fig:stepu}
\end{center}
\end{figure}

\begin{figure}[hbt!]
\begin{center}
\includegraphics[height=1in,angle=0,width=0.8\linewidth]{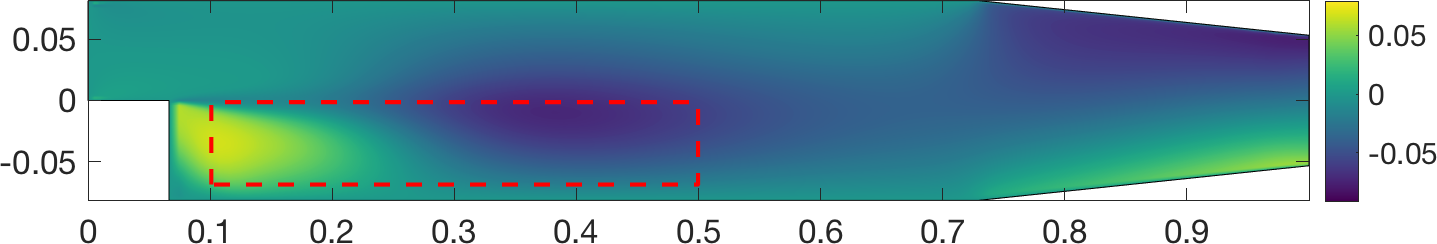} \\[1em] 
\includegraphics[height=1in,angle=0,width=0.8\linewidth]{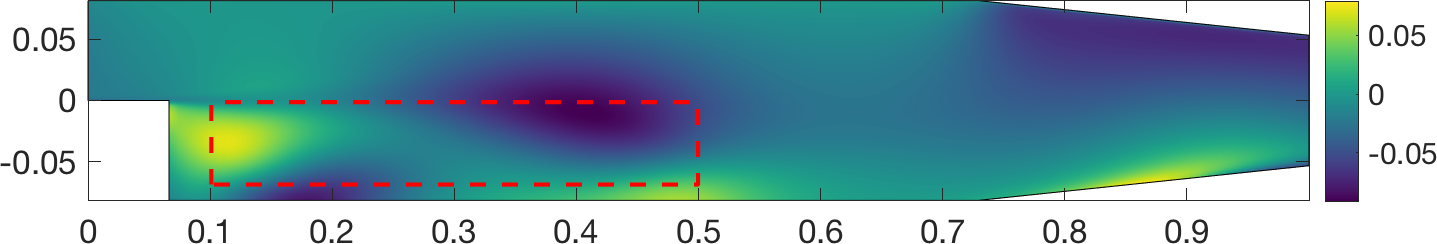} 
\captionof{figure}{Contours of the velocity component $v$ in the vertical direction for the steady turbulent flow over a backward facing step. (Top) from the {\it SST} $k -\omega$ model; (Bottom) inferred by the physics-informed neural network. The dash-lined box indicates the region where the flow information is missing.}
\label{fig:stepv}
\end{center}
\end{figure}

We first consider a steady turbulent flow over a backward facing step, experimentally studied by Pitz and Daily date back to 1981 \cite{pitz1981}. The Reynolds number is $5\times 10^5$ based on the step height. The {\it SST} $k -\omega$ model was adopted to generate the data. Here SST denotes ``{\it shear-stress transport}'' \cite{spalar2009}. The contours of velocity components along the streamwise and vertical directions are given in Figs. \ref{fig:stepu}(a) and \ref{fig:stepv}(a), respectively. It is assumed that in the dash-lined rectangular box, denoted by $\Omega^\prime$, the flow information  therein is missing and thus not used as training data. 

\begin{figure}[hbt!]
\begin{center}
\includegraphics[height=0.94in,angle=0,width=0.8\linewidth]{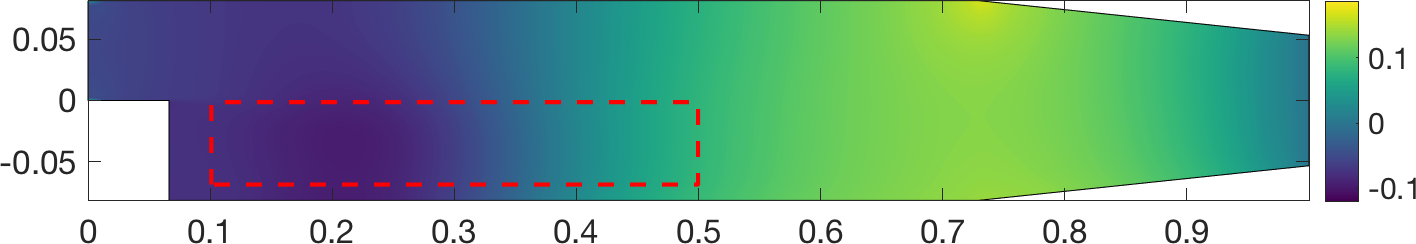} \\[1em]
\includegraphics[height=0.94in,angle=0,width=0.8\linewidth]{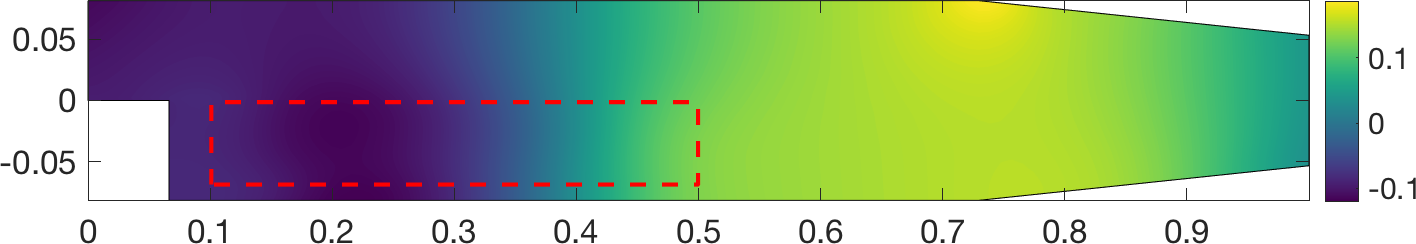}
\captionof{figure}{The pressure fields for the steady turbulent flow over a backward facing step. (Top) from the {\it SST} $k-\omega$ model; (Bottom) inferred by the physics-informed neural network. }
\label{fig:stepp}
\end{center}
\end{figure}

\begin{figure}[hbt!]
\begin{center}
\includegraphics[height=1.1in,angle=0,width=0.8\linewidth]{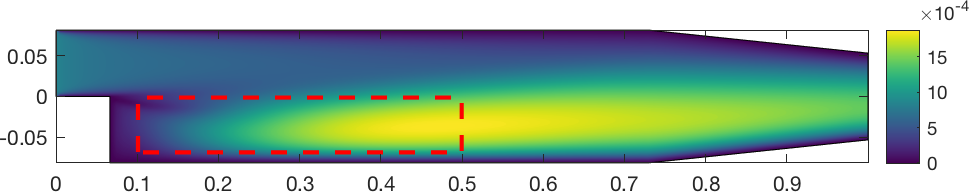} \\
\includegraphics[height=1.1in,angle=0,width=0.8\linewidth]{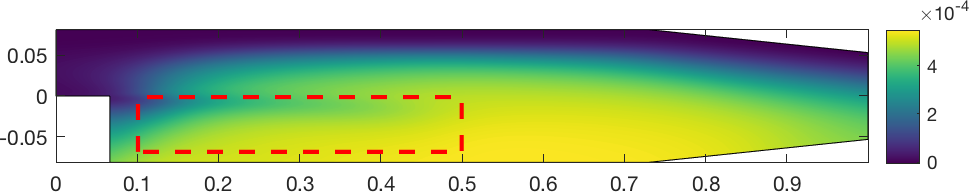} 
\captionof{figure}{Comparison between the eddy viscosity and the artificial viscosity for the steady turbulent flow over a backward facing step. (Top) eddy viscosity $\nu_t$ from the {\it SST} $k -\omega$ model; (Bottom) artificial viscosity $\nu_{\rm \eta}$ inferred by the physics-informed neural network.}
\label{fig:stepnut}
\end{center}
\end{figure}

After the neural network is trained with the data outside $\Omega^\prime$, together with constrains of the governing equations, the flow variables in the whole domain $\Omega$ are inferred. In Figs. \ref{fig:stepu}(b) and \ref{fig:stepv}(b), the inferred velocity components by the physics-informed deep learning are given.  It is clear that in the box region, we  get the similar  velocity fields as illustrated in Figs. \ref{fig:stepu}(a) and \ref{fig:stepv}(a), demonstrating that the averaged flow dynamics is mimicked correctly with the proposed parameterised NSEs. In Fig. \ref{fig:stepp}, the inferred pressure field is in good agreement with the reference one from the {\it SST} $k-\omega$ model. It is interesting, on the other hand, to compare the inferred artificial viscosity $\nu_{\rm \eta}$ with eddy viscosity $\nu_t$ empirically determined by the {\it SST} $k-\omega$ model. Although they have different meaning thus are different in both magnitude and spatial distribution, as shown in Fig. \ref{fig:stepnut}, their distributions are related to the flow structure and both are relatively large in the region with separation bubble. 

\begin{figure}[hbt!]
\begin{center}
\includegraphics[height=0.94in,angle=0,width=0.8\linewidth]{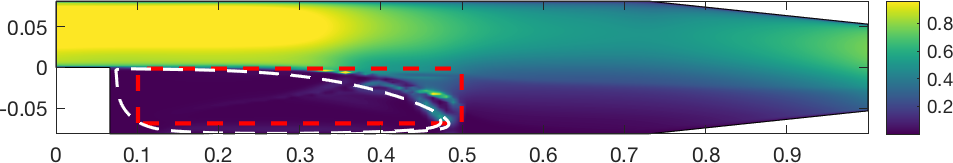} \\
\includegraphics[height=0.94in,angle=0,width=0.8\linewidth]{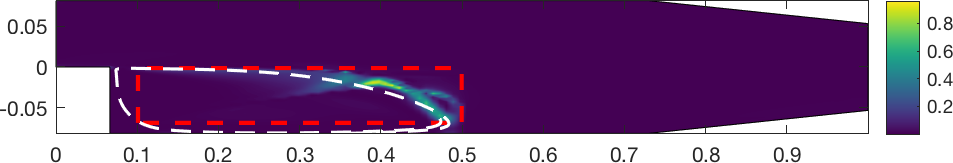} 
\captionof{figure}{The total energy and `{\it perturbation}' energy of the inferred velocity for the steady turbulent flow over a backward facing step. (Top) the inferred energy with the original NSEs; (Bottom) the `{\it perturbation}' energy. The white dashed-line indicates the separation region of the averaged field, while the red dash-lined box is the region where the flow information is missing.}
\label{fig:stepenergy}
\end{center}
\end{figure}

For the same steady turbulent flow, we specially consider a case where the parameter $\nu_{\rm \eta}$ is removed and thus the original NSEs are used as a physics constraint during the learning. The energy contours of the inferred velocity field $\mathbf{u}_{\rm NSEs}$ and the `{\it perturbation}' field ($\mathbf{u}_{\rm NSEs}-\bar{\mathbf{u}}$) are given in Fig. \ref{fig:stepenergy}. Special structures are observed around the tip of the separation bubble.  Compared with the solution based on the parameterised NSEs, this significant inconsistency provides some potential information about the region where the most unstable modes or the energetic turbulent structures emerge around the flow separation region.

As a further demonstration,  we consider the time-dependent turbulent flow past a cylinder. The Reynolds number is, $Re = u_{in} D/\nu = 10^5$, where $u_{in}$ and $D$ are the inlet velocity and diameter of the cylinder, respectively. The reference flows are generated respectively by the {\it SST} $k -\omega$ and {\it SST-SAS} turbulence models. It should be emphasized that, unlike other traditional RANS models (e.g. the {\it SST} $k -\omega$ model), the {\it 
SST-SAS} model has a pure RANS nature but achieves LES behavior. With this 
time-dependent turbulent flow, we demonstrate that the proposed methodology has a great adaptability to turbulence models for unsteady flow dynamics. In practice, the reference flow data obtained by the {\it SST-SAS} and {\it SST} $k -\omega$ turbulence models can be regarded as the ``{\it filtered}" experimental data. 

\begin{figure}[hbt!]
\begin{center}
\includegraphics[height=4in,angle=0,width=0.8\linewidth]{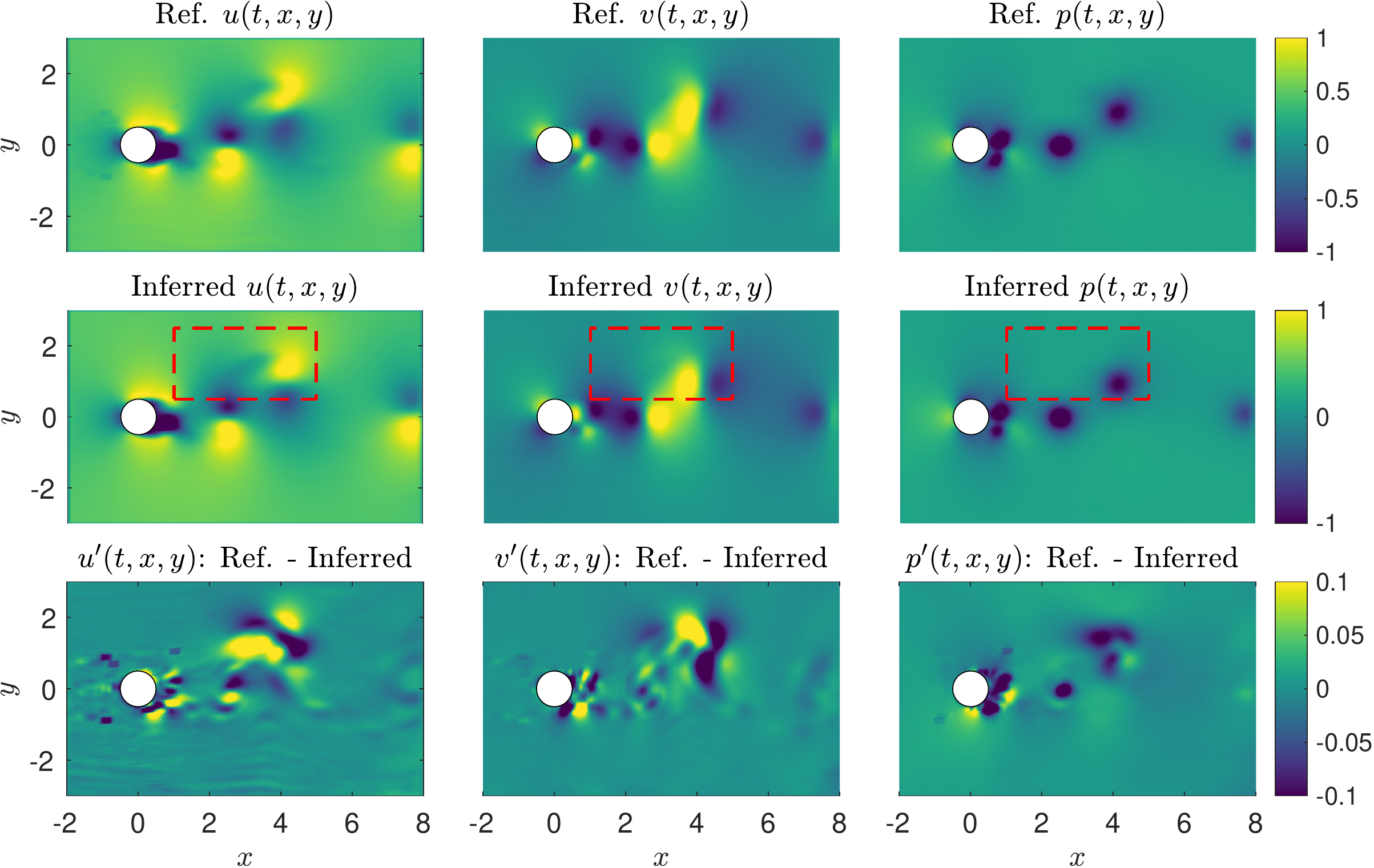} 
\captionof{figure}{Unsteady turbulent flow past a cylinder: (Top) the instantaneous reference velocity and pressure fields at $t=11.96$, generated by the {\it SST-SAS} turbulence model; (Mid) the inferred fields at the same time instant with the missing dynamics; (Bottom) the difference between the reference fields and the inferred fields. The red dashed-line indicates the region where the flow dynamics is missing. }
\label{fig:sas}
\end{center}
\end{figure}

\begin{figure}[hbt!]
\begin{center}
\includegraphics[height=2.5in,angle=0,width=0.8\linewidth]{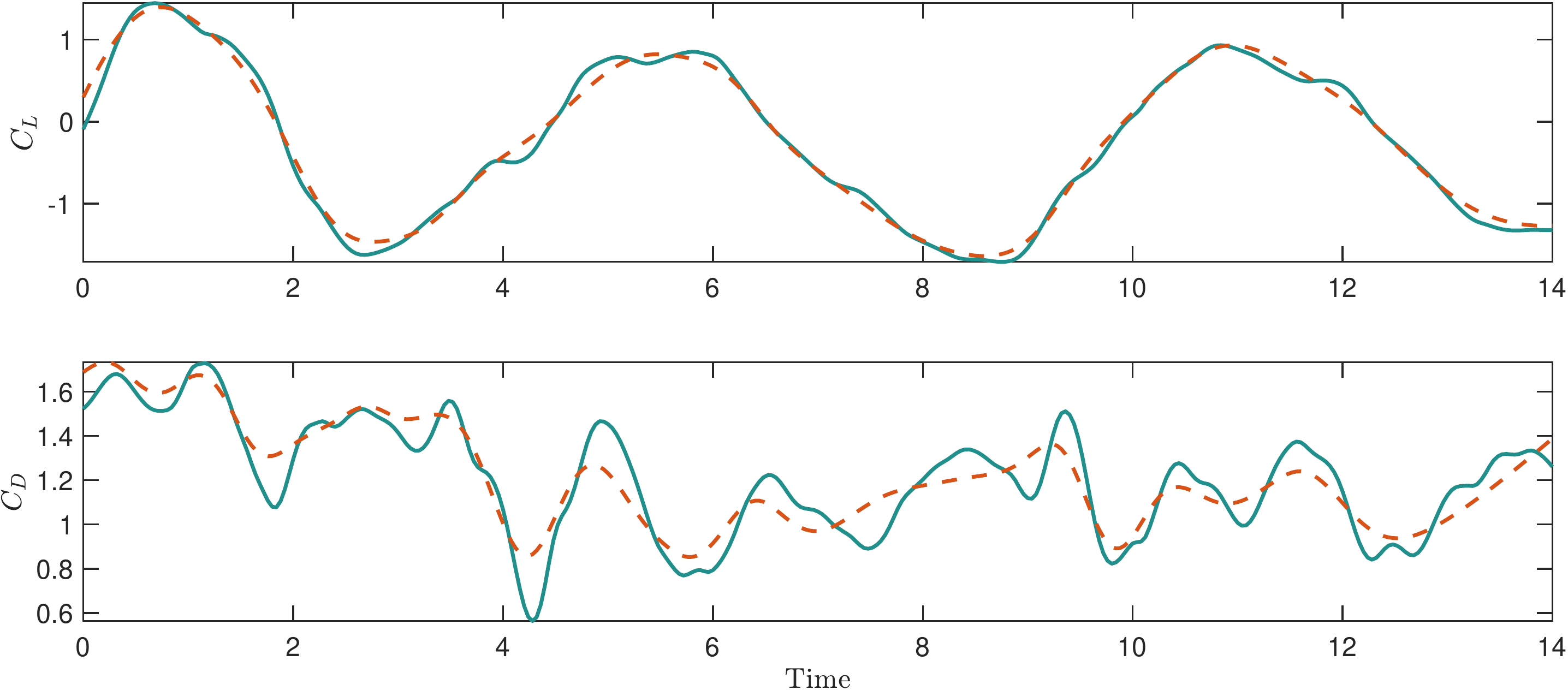} 
\captionof{figure}{Lift and drag coefficients over the cylinder: (Top) profiles of the lift coefficient; (Bottom) profiles of the drag coefficient. The solid curves are generated from the reference data by the {\it SST-SAS} turbulence model, while the dashed curves generated from the inferred fields.}
\label{fig:ld_sas}
\end{center}
\end{figure}

\begin{figure}[hbt!]
\begin{center}
\includegraphics[height=1.8in,angle=0,width=0.8\linewidth]{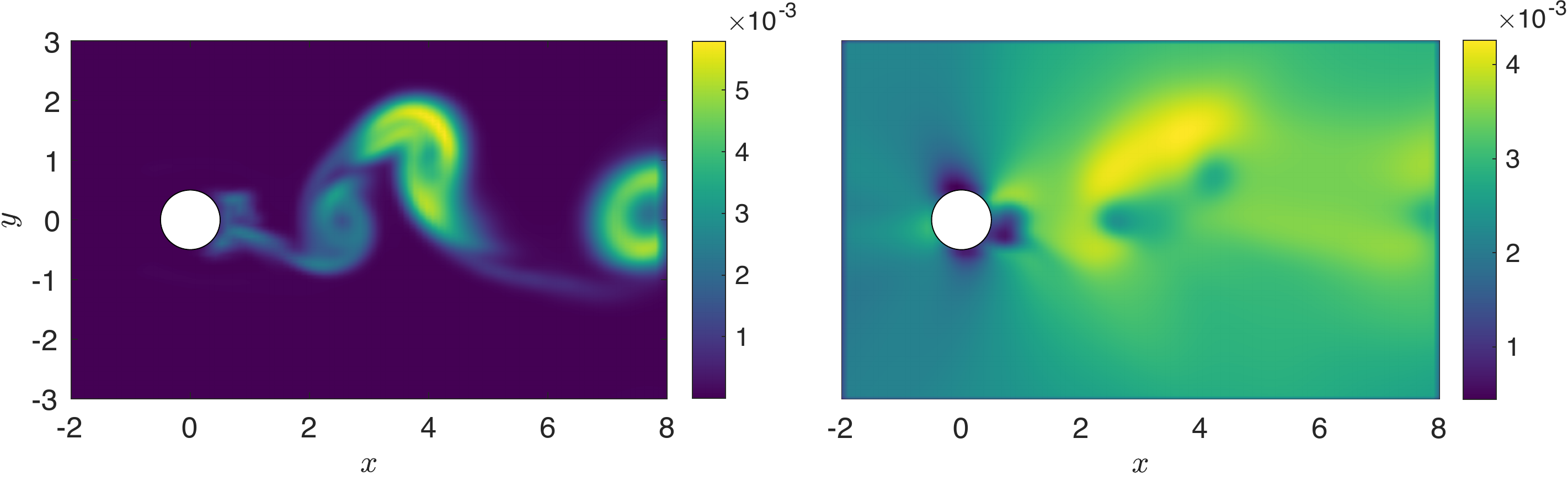} 
\captionof{figure}{Comparison between the eddy viscosity and the artificial viscosity for the unsteady turbulent flow past a cylinder: (Left) eddy viscosity $\nu_t$ from {\it SST-SAS} turbulence model; (Right) the parameter $\nu_{\rm \eta}$ inferred by the physics-informed neural network.}
\label{fig:nut_sas}
\end{center}
\end{figure}

It is assumed, again, that in the system there exists a local region where the flow dynamics is missing, as indicated by a dashed rectangular box in Fig. \ref{fig:sas}.  In order to employ the established methodology, we consider a series of flow field snapshots obtained by the {\it 
SST-SAS} model, and only a part of the velocity field beyond the dashed-rectangular box on every snapshot is employed for training. In Fig. \ref{fig:sas}(Mid), the reconstruction 
of the missing flow dynamics at $t=11.96$ in the dashed-rectangular box is presented. Surprisingly, compared with Fig. \ref{fig:sas}(Top),  the reconstructed missing flow  coincides with the reference flow  very well, even with complex vortex structure in the box. Qualitatively, we show the differences between the reference fields and the inferred ones in Fig. \ref{fig:sas}(Bottom). Relative low deviation is observed. Meanwhile, the profiles of the lift and drag coefficients are given in Fig. \ref{fig:ld_sas}. The inferred $C_L$ profile is consistent with the reference one.  However, there exist some differences between the inferred $C_D$ and the reference  one. As observed from the comparison, the unsteadiness (or ``{\it irregularity}'' ) of the $C_D$ profile from the inferred data is reconstructed fairly well. Further, we show the distribution of the viscosity in Fig. \ref{fig:nut_sas}. It is observed that the patterns of the viscosity from both the {\it SST-SAS} and neural network, are consistent with the vortex shedding pattern, but with different magnitudes. 

\begin{figure}
\begin{center}
\includegraphics[height=4in,angle=0,width=0.8\linewidth]{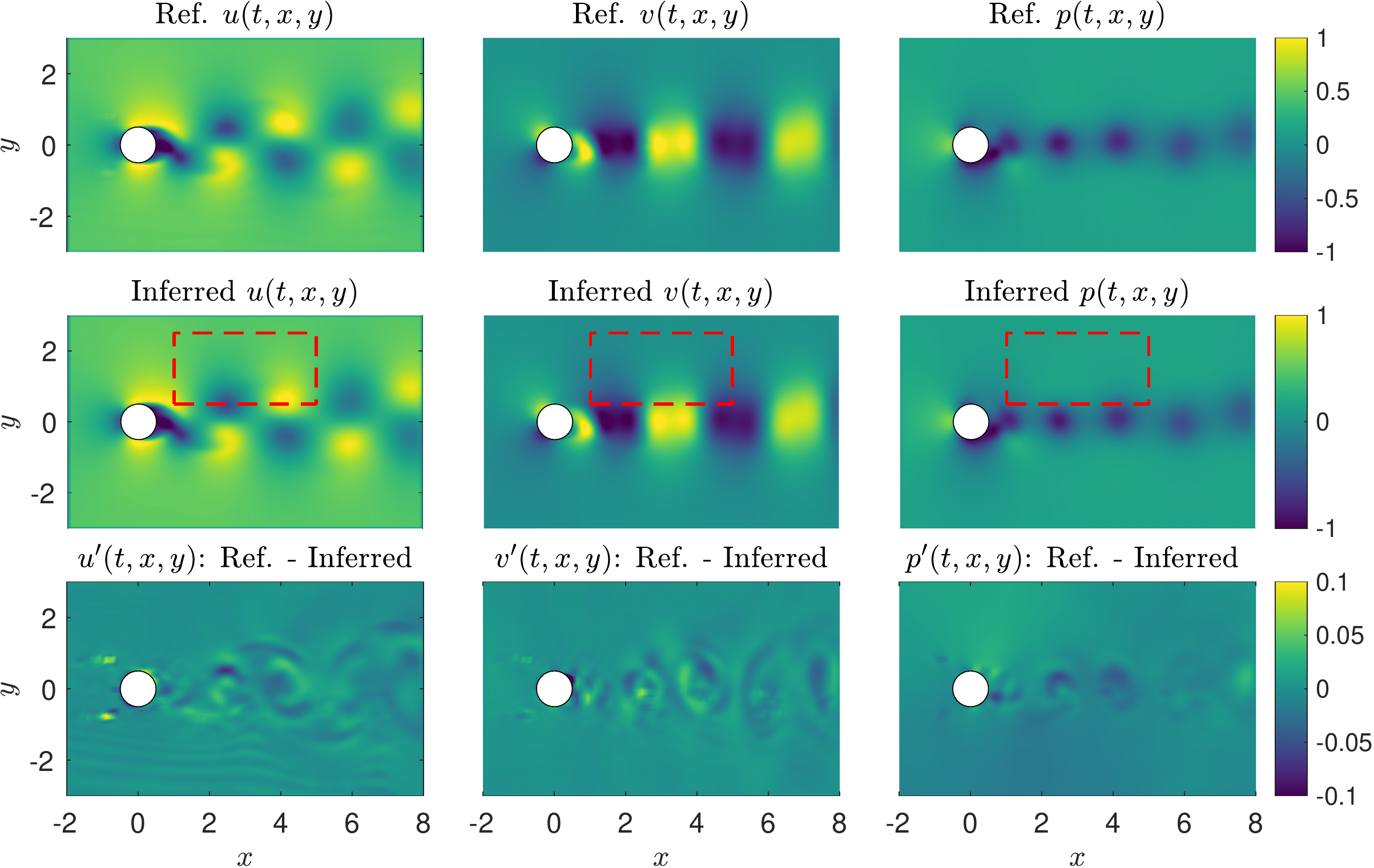} 
\captionof{figure}{Unsteady turbulent flow past a cylinder: (Top) the instantaneous reference velocity and pressure fields at $t=7$, generated by the {\it SST} $k -\omega$ turbulence model; (Mid) the inferred fields at the same time instant with the missing dynamics; (Bottom) the difference between the reference fields and the inferred fields. The red dashed-line indicates the region where the flow dynamics is missing.}
\label{fig:sst}
\end{center}
\end{figure}

\begin{figure}
\begin{center}
\includegraphics[height=2.5in,angle=0,width=0.8\linewidth]{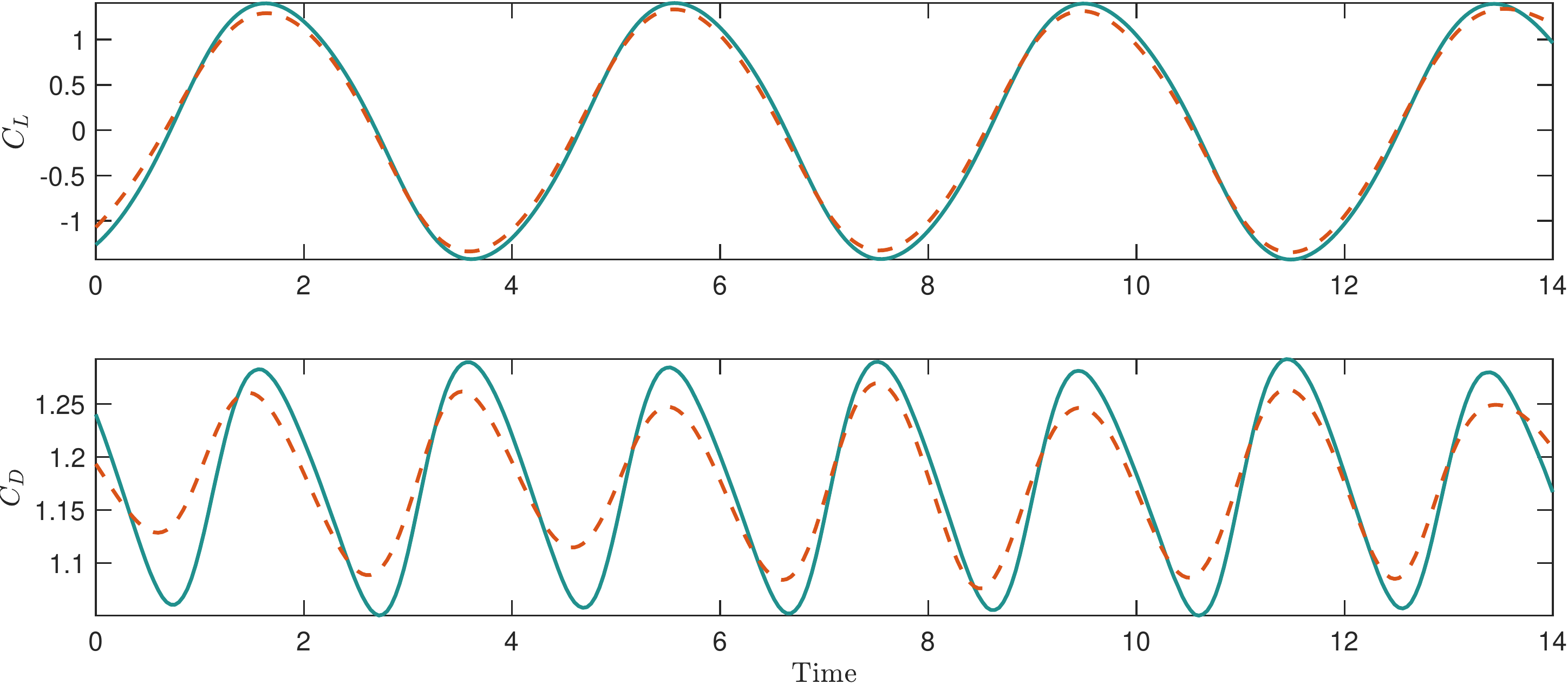} 
\captionof{figure}{Lift and drag coefficients over the cylinder: (Top) profiles of the lift coefficient; (Bottom) profiles of the drag coefficient. The solid curves are generated from the reference case by the the {\it SST} $k -\omega$ turbulence model, while the dashed curves generated from the inferred fields.}
\label{fig:ld_sst}
\end{center}
\end{figure}

\begin{figure}
\begin{center}
\includegraphics[height=1.8in,angle=0,width=0.8\linewidth]{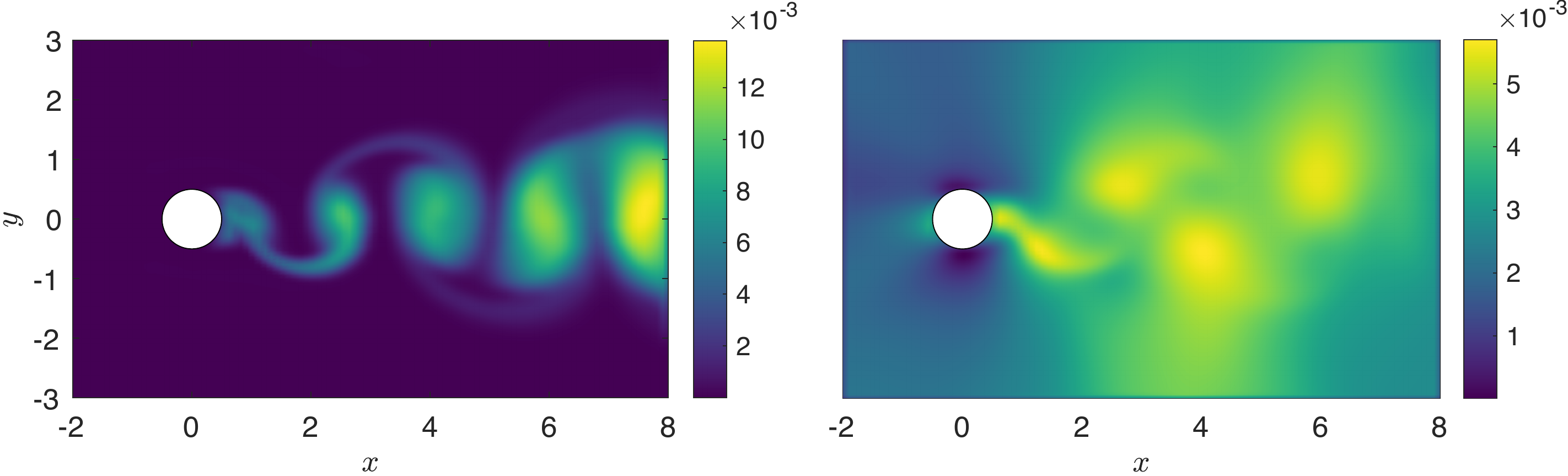}
\captionof{figure}{Comparison between the eddy viscosity and the artificial viscosity for the unsteady turbulent flow past a cylinder: (Left) eddy viscosity $\nu_t$ from {\it SST} $k -\omega$ turbulence model; (Right) the parameter $\nu_{\rm \eta}$ inferred by the physics-informed neural network.}
\label{fig:nut_sst}
\end{center}
\end{figure}

Accordingly, for the flow past a cylinder, instantaneous results at $t=7$ with the {\it SST} $k -\omega$ model are given in Figs. \ref{fig:sst}-\ref{fig:nut_sst}. We see that using the methodology with the physics-informed deep learning, the missing flow dynamics 
whose location is assigned arbitrarily, is recovered properly with good accuracy. 

With the cases investigated, we demonstrate that when the parameter $\nu_{\rm \eta}$ is introduced as an output variable of the neural network, the parameterised governing system of flow dynamics has a strong adaptivity to approximate the data from the turbulence models. Significantly, the missing data in a randomly specified region can be inferred by the physics-informed deep learning. Given the reality that reasonable accurate velocity and pressure are predicted by the methodology, the inferred $\nu_{\rm \eta}$ indicates that the eddy viscosity in the conventional turbulence models probably has an over-dissipation effect on the flow structures and a different spatial distribution from the inferred one. This discovery implies that the physics-preserved data driven techniques may have a transformative influence on the turbulence closure modelling.

\section{Conclusion and Discussion}
The methodology proposed in this paper provides a way to explore the missing flow dynamics, which is well validated by the data from the turbulence models. We show that from the single parameterisation of the Navier-Stokes equations, the feasibility and potential of the method are significant for the ``{\it averaged}'' or ``{\it filtered}'' data, no matter obtained experimentally or numerically. Surprisingly, the current methodology allows to use the parameterised governing equations without extra modellings, which is shaping a perspective in exploring the dynamics beyond the description of the original governing system, although the deviation of exploring missing dynamics exists. From this point of view, the parameterisation, in fact, provides a solid foundation of accurately describing practical complex flow dynamics, which may not be accurately described by the traditional turbulence modellings where the artificial viscosity $\nu_{\rm t}$ needs to be further tuned.

The proposed methodology provides a new way for further calibrations of the turbulence closures, which is probably difficult for traditional case-by-case machine learning with simply tuning the scalar modelling parameters. As indicated from the theory of turbulence modelling, due to the significance of the parameter $\nu_{\rm \eta}$, the proposed method is naturally termed as ``{\it data-driven scale-adaptive turbulent structure recovering}''.

Although only 2D incompressible flows were considered as examples in this work, the proposed methodology can be extended to 3D incompressible or compressible flows. While learning from the parameterised Navier-Stokes equations is successful in coping with the traditional fluid dynamics problems, this, in turn, implies generalisation of the employed methodology could enable exploration and exploitation in a broad range of complex systems, e.g. weather prediction, ocean recirculation and biofluid mechanics.



\section*{Acknowledgments}
The authors would like to acknowledge support from National Numerical Wind Tunnel Project (Grant NO. NNW2019ZT4-B09) and the National Natural Science Foundation of China (Grant NOs. 91852106).

\section*{DATA AVAILABILITY}

The data that support the findings of this study are available from the corresponding author upon reasonable request.

\section*{references}
\bibliographystyle{plainnat}
\bibliography{PoF}

\end{document}